\documentclass[12pt,a4paper]{article}
\usepackage{epsfig}
\begin{document}
\title{Effects of the littlest Higgs model with T-parity on Higgs boson production at high energy $e^{+}e^{-}$ colliders }

\author{Chong-Xing Yue and Nan Zhang \\
{\small  Department of Physics, Liaoning Normal University, Dalian
116029, China}\thanks{E-mail:cxyue@lnnu.edu.cn}\\}
\date{\today}

\maketitle
\begin{abstract}
 The Higgs boson  production processes $e^{+}e^{-}\rightarrow
 ZH$, $e^{+}e^{-}\rightarrow \bar{\nu_{e}}\nu_{e}H$, and
 $e^{+}e^{-}\rightarrow t\bar{t}H$ are very important for
 studying Higgs boson properties and further testing new physics
 beyond the standard model($SM$) in the high energy linear
 $e^{+}e^{-}$ collider($ILC$). We estimate the contributions of the
 littlest Higgs model with T-parity($LHT$ model) to these processes
 and find that the $LHT$ model can generate significantly
 corrections to the production cross sections of these processes.
 We expect the possible signals of the $LHT$ model can be
 detected via these processes in the future $ILC$ experiments.

\end{abstract}

\newpage

The discovery and study of Higgs boson is one of the most important
goals of present and future high energy collider experiments[1]. The
physics of the Higgs boson will be explored by high energy
experiments at the upgraded Tevatron collider at Fermlab and in the
near future at the large hadron collider at $CERN$($LHC$). The $LHC$
will make the first exploration of the TeV energy range, and will be
able to discover Higgs boson in the full mass range, provided it
exists[2]. After discovery of the Higgs boson at the $LHC$, one of
the most pressing tasks is a proper determination of its properties
 since it is very important to study the mechanism of
electroweak symmetry breaking($EWSB$), the underlying origin of
mass, and test the new physics beyond the standard model ($SM$). The
$LHC$ will be able to finish a few measurement on the couplings of
the Higgs boson to fermions and gauge bosons but the most precise
measurements will be performed in the clean environment of a future
high energy linear $e^{+}e^{-}$ collider($ILC$)[3]. Furthermore,
discoveries at the $LHC$ may also point to physics scales beyond the
reach of the $ILC$, this area could be accessed later by a multi-TeV
$e^{+}e^{-}$collider[4].

The main production mechanism of the Higgs boson $H$ at the $ILC$
are the Higgs-strahlung process $e^{+}e^{-}\rightarrow ZH$[5] and
the $WW$ fusion process $e^{+}e^{-}\rightarrow
\bar{\nu_{e}}\nu_{e}W^{*}W^{*} \rightarrow\bar{\nu_{e}}\nu_{e}
H$[6]. The cross section for the Higgs-strahlung process decreases
as $s^{-1}$($\sqrt{s}$ is the center-of-mess energy) and dominates
at low energies, while the cross section for the $WW$ fusion process
rise as $log(s/m^{2}_{H})$ and dominates at high energies. The $ZZ$
fusion process $e^{+}e^{-}\rightarrow
e^{+}e^{-}Z^{*}Z^{*}\rightarrow e^{+}e^{-}H $ can also contribute to
the Higgs boson production. However, the cross section is suppressed
by an order of magnitude compared to that for the $WW$ fusion
process, due to the ratio of the $W^{\pm}e\nu_{e}$ coupling to the
$Ze\bar{e}$ coupling, 4$C^{2}_{W}\simeq3$. From the production cross
sections for the Higgs-strahlung and $WW$ fusion processes, the
absolute values of the Higgs couplings to the electroweak gauge
bosons $Z$ and $W $, and also the Higgs couplings to quark and
leptons can be determined to a few percent in a model independent
way[1,3].

The top quark, with a mass of the order of the electroweak scale, is
the heaviest particle yet discovered. The coupling of Higgs boson to
top quark pair, which is the largest one among the Yukawa couplings,
should play a key role in a theory generating fermion masses and is
particularly sensitive to the underlying physics. Thus, studying the
Yukawa coupling $Ht\bar{t}$ is of particular interest. For a light
Higgs boson, precise determination of this coupling can be performed
at the $ILC$ via the associated production process
$e^{+}e^{-}\rightarrow t\bar{t}H$[7]. A precision of around 5\% can
be reached at an $ILC$ with $\sqrt{s}$=800$\sim$1000GeV and the
integrated luminosity ${\cal L}_{int}\simeq1000fb^{-1}$[3,4].

The aim of this letter is to consider the processes
$e^{+}e^{-}\rightarrow ZH$, $e^{+}e^{-}\rightarrow
\bar{\nu_{e}}\nu_{e}H$, and $e^{+}e^{-}\rightarrow t\bar{t}H$ in the
context of the littlest Higgs model with T-party i. e. $LHT$
model[8], and see whether the effects of this model on these
processes can be detected in the future $ILC$ experiments.

To solve the so-called hierarchy or fine-tuning problem of the $SM$,
the little Higgs models were proposed as kind of $EWSB$ mechanism
accomplished by a naturally light Higgs sector[9]. The key feature
of this kind of models is that the Higgs boson is a pseudo-Goldstone
boson of a global symmetry which is spontaneously broken at some
higher scale $f$. So far, a number of specific models have been
proposed, which differ in the assumed higher symmetry and in the
representation of the scalar multiplets. The littlest Higgs($LH$)
model[10] is one of the simplest and phenomenologically viable
models, which realizes the little Higgs idea. Most phenomenological
analysis about the little Higgs models are given in the context of
the $LH$ model[11,12].

It has been shown that the $LH$ model suffers from severe
constraints from precision electroweak measurement, which could only
be satisfied by fine-tuning the model parameters[11,13]. To avoid
this problem, T-parity is introduced into the $LH$ model, which is
called $LHT$ model[8]. In the $LHT$ model, all the $SM$ particles
are assigned with an even T-parity, while all the new particles are
assigned with an odd T-parity, except for the little Higgs partner
of the top quark. Thus, the $SM$ gauge bosons can not mix with the
new gauge bosons, and the electroweak precision observables are not
modified at tree level. It has been shown that loop corrections to
precision electroweak obserbables are much small and the scale
parameter $f$ can be decreased to 500GeV[8,14]. Thus, the $LHT$
model can produce rich phenomenology in the present and future high
energy experiments[15,16,17,18].

In the $LHT$ model, all the $SM$ particles are T-even, while most of
the new particles appeared at the TeV scale are T-odd, except for
the heavy vector-like top quark T$^{+}$. Thus, the couplings of the
electroweak gauge bosons to light fermions are not modified from
their corresponding $SM$ couplings at tree level. However, since the
new T-odd fermions and T-even heavy top quark T$^{+}$ are introduced
into the $LHT$ model, the couplings of the electroweak gauge bosons
to top quark and the couplings of the Higgs bosons to ordinary
particles are corrected at tree level [15,16]. The expressions of
the couplings $HVV(V=W$ or $Z)$, $Ht\bar{t}$,  $Zt\bar{t}$, and
$ZtT^{+}$, which are related to the processes $e^{+}e^{-}\rightarrow
\bar{\nu_{e}}\nu_{e}H$, $e^{+}e^{-}\rightarrow ZH$, and
$e^{+}e^{-}\rightarrow t\bar{t}H$, can be written as:

\begin{eqnarray}
g_{HV_{\mu}V_{\nu}}&=&\frac{2M^{2}_{V}}{\nu}[1-\frac{1}{4}\frac{\nu^{2}}{f^{2}}-\frac{1}{32}\frac{\nu^{4}}
{f^{4}}]g_{\mu\nu}
;\\
g_{Ht\bar{t}}&=&-\frac{m_{t}}{\nu}[1-(\frac{3}{4}-C_{\lambda}^{2}+C_{\lambda}^{4})\frac{\nu^{2}}{f^{2}}]
;\\
g^{L}_{Zt\bar{t}}&=&\frac{e}{S_{W}C_{W}}[\frac{1}{2}-\frac{2}{3}S_{W}^{2}-\frac{C_{\lambda}^{4}}{2}
\frac{\nu^{2}}{f^{2}}],\hspace{0.5cm}
 g^{R}_{Z\bar{t}t}=-\frac{e}{S_{W}C_{W}}\frac{2}{3}S_{W}^{2};\\
  g^{L}_{Ht\bar{T}^{+}}&=&-\frac{m_{t}}{f}S_{\lambda}^{2},\hspace{0.5cm}
g^{R}_{Ht\bar{T}^{+}}=\frac{m_{t}}{\nu}\frac{C_{\lambda}}{S_{\lambda}};\\
g^{L}_{Zt\bar{T}^{+}}&=&-\frac{e}{2S_{w}C_{w}}C_{w}^{2}\frac{\nu}{f}\sqrt{1-C_{\lambda}^{4}\frac
{\nu^{2}}{f^{2}}},\hspace{0.5cm}g^{R}_{Zt\bar{T}^{+}}=0.
\end{eqnarray}
Where $\nu\simeq$246 is the electroweak scale,
$S_{W}=Sin\theta_{W}$, $\theta_{W}$ is the Weinberg angle, $f$ is
the scale parameter. The mixing parameter
$C^{2}_{\lambda}=\frac{\lambda^{2}_{1}}{\lambda^{2}_{1}+\lambda^{2}_{2}}$($S_{\lambda}^{2}
=1-C^{2}_{\lambda}$), in which $\lambda_{1}$ and $\lambda_{2}$ are
the Yukawa coupling parameters.
\begin{figure}[htb]
\begin{center}
\epsfig{file=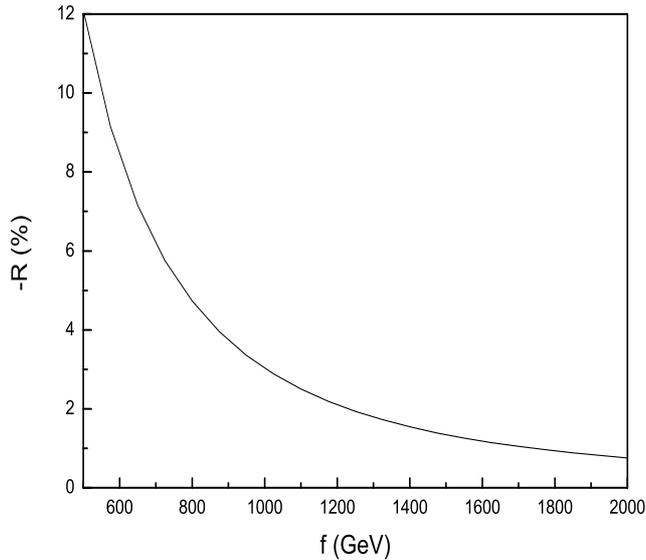,width=270pt,height=250pt} \vspace{-1.0cm}
 \caption{The relative correction parameter $R$ as a function of the scale
parameter $f$.} \label{ee}
\end{center}
\end{figure}

From above discussions, we can see that the $LHT$ model can generate
corrections to the production cross sections for the processes
$e^{+}e^{-}\rightarrow \bar{\nu_{e}}\nu_{e}H$ and
$e^{+}e^{-}\rightarrow ZH$ via the modification of the couplings
$HWW$ and $HZZ$. The value of the relative correction parameters
$R_{1}=\frac{\sigma^{LHT}(H)-\sigma^{SM}(H)}{\sigma^{SM}(H)}$ and
$R_{2}=\frac{\sigma^{LHT}(ZH)-\sigma^{SM}(ZH)}{\sigma^{SM}(ZH)}$ are
easily calculated, in which $\sigma^{SM}(H)$ and $\sigma^{SM}(ZH)$
represent the corresponding cross sections predicted by the $SM$.
From Eq.(1), we can see that there is $R_{1}=R_{2}$, thus we have
taken $R=R_{1}=R_{2}$ and plotted $R$ as a function of the scale
parameter $f$ in Fig.1.

For the $LH$ model, the extra contributions to the process
$e^{+}e^{-}\rightarrow \bar{\nu_{e}}\nu_{e}H$ come from the heavy
gauge bosons $W^{\pm}_{H}$, the modification of the relations among
the $SM$ parameters and the precision electroweak input parameters,
and the correction terms of the $SM$ $W_{e}\nu_{e}$ and $HWW$
couplings[19]. Considering the constraints of the electroweak
precision data on the $LH$ model, the scale parameter $f$ should be
in the range of 1TeV$\sim$3TeV[13]. If we take $f$$\geq$2TeV, the
relative correction of the $LH$ model to the production cross
section for the process $e^{+}e^{-}\rightarrow
\bar{\nu_{e}}\nu_{e}H$ is smaller than 5\%. Thus, the effects of the
$LHT$ model on this process are easy to be detected than those of
the $LH$ model in the future $ILC$ experiments. The contributions of
the $LH$ model to the process $e^{+}e^{-}\rightarrow ZH$ mainly come
from the heavy photon $B_{H}$ [20]. For reasonable values of the
mixing parameter $c'$ and the mass parameter $M_{B_{H}}$, the
relative corrections of the $LH$ model to the production cross
section for the process $e^{+}e^{-}\rightarrow ZH$ can be
significantly large. Thus, the contributions of the $LHT$ model to
this process can be larger or smaller than those of the $LH$ model.

In the $SM$, the process $e^{+}e^{-}\rightarrow t\bar{t}H$ proceeds
mainly through Higgs boson emission off top quarks, while emission
from intermediate $Z$ boson plays only a minion role, which has been
extensively studied in Ref.[7]. The contributions of the $LH$ model
to this process mainly come from the new gauge bosons $B_{H}$ and
$Z_{H}$. In sizable regions of the parameter space preferred by the
electroweak precision data, the absolute value of the relative
correction parameter
$R_{3}=\delta\sigma(t\bar{t}H)/\sigma^{SM}(t\bar{t}H)$ can be larger
than 5\%[21]. For the $LHT$ model, the T-odd new gauge bosons can
not contribute to this process at tree level. Thus, the
contributions of the $LHT$ model to the production cross section of
the process $e^{+}e^{-}\rightarrow t\bar{t}H$ only come from the
heavy T-even top-quark T$^{+}$ and the correction terms to the $SM$
couplings $Zt\bar{t}, Ht\bar{t},$ and $HZZ$.

\begin{figure}[htb]
\begin{center}
\epsfig{file=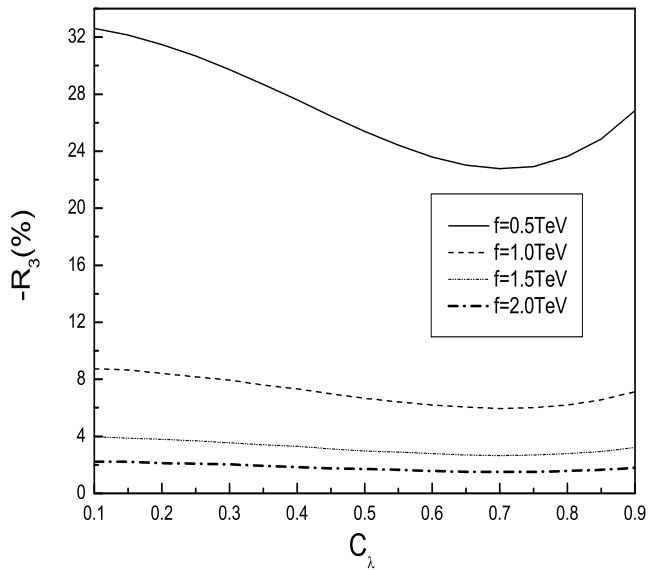,width=270pt,height=250pt} \vspace{-1.0cm}
 \caption{The relative correction parameter $R_{3}$ as a function of the mixing parameter $C_{\lambda}$
\hspace*{1.8cm}for four values of the scale  parameter $f$.}
\label{ee}
\end{center}
\end{figure}

Similar to Ref.[21], we can give the production cross section of the
process $e^{+}e^{-}\rightarrow t\bar{t}H$ in the context of the
$LHT$ model. The relative correction parameter $R_{3}$ generated by
the $LHT$ model is plotted as a function of the mixing parameter
$C_{\lambda}$ for four values of the scale parameter $f$ in $Fig.2$.
One can see from $Fig.2$ that, as long as $f\leq$1TeV, the absolute
value of $R_{3}$ is larger than 7\%, which might be detected in the
future $ILC$ experiments.

Since the scale parameter $f$ of the $LHT$ model can be allowed to
be lower than 1TeV, this model can produce rich phenomenology in the
present and future experiments. In this letter, we estimate the
corrections of the $LHT$ model to the production cross sections of
the processes $e^{+}e^{-}\rightarrow ZH$, $e^{+}e^{-}\rightarrow
\bar{\nu_{e}}\nu_{e}H$, and $e^{+}e^{-}\rightarrow t\bar{t}H$, which
are very important for studying Higgs properties and test new
physics beyond the $SM$. Our numerical results show that, with
reasonable values of the free parameters, the $LHT$ model can
generate significantly contributions to these processes, which might
approach the observable threshold of the near-future $ILC$
experiments.

\vspace{0.5cm} \noindent{\bf Acknowledgments}

C. X. Yue would like to thank the {\bf Abdus Salam } International
Centre for Theoretical Physics(ICTP) for partial support. This work
was supported in part by Program for New Century Excellent Talents
in University(NCET-04-0290), the National Natural Science Foundation
of China under the Grants No.10475037 and 10675057.

\null
\end{document}